\documentclass[aps,pra,twocolumn,showpacs,epsfig]{revtex4-1}
\usepackage{color,graphicx}
\usepackage{amsmath,amssymb,amsthm,dsfont}

\usepackage{epsfig}
\usepackage{epstopdf}
\usepackage{color}
\newcommand{\ket}[1]{\left\vert#1\right\rangle}

\newcommand{\nbar}{\overline{n}}

\begin{document}

\title{Tuning non-Markovianity by spin-dynamics control}

\author{Salvatore Lorenzo$^{1,2}$, Francesco Plastina$^{1,2}$, and Mauro Paternostro$^{3}$}
\affiliation{$^1${Dipartimento di  Fisica, Universit\`a della Calabria, 87036 Arcavacata di Rende (CS), Italy}\\
$^2$ INFN - Gruppo collegato di Cosenza \\
$^3$ Centre for Theoretical Atomic, Molecular and Optical Physics, School of Mathematics and
  Physics, Queen's University, Belfast BT7 1NN, United Kingdom}
\date{\today}

\begin{abstract}
We study the interplay between forgetful and memory-keeping
evolution enforced on a two-level system by a multi-spin
environment whose elements are coupled to local bosonic
baths. Contrarily to the expectation that any non-Markovian effect
would be {\it buried} by the forgetful mechanism induced by the
spin-bath coupling, one can actually induce a full
Markovian-to-non-Markovian transition of the two-level system's
dynamics, controllable by parameters such as the mismatch between the energy of the  two-level system and of the spin
environment. For a symmetric coupling,
the amount of non-Markovianity surprisingly grows with the
number of decoherence channels.
\end{abstract}
 \maketitle

\section{Introduction}
The understanding of the implications of non-Markovianity
and the reasons for its occurrence are still largely elusive. Yet,
they are stimulating a growing interest in light of their
potential impact on many disciplines, from quantum information and
 nano-technology up to quantum
biology~\cite{hu,qbio,Rebentrost}. An important contribution to
this quest  came from the formulation of quantitative measures of
the degree of non-Markovianity of a process~\cite{Wolf,Huelga,BreuerMea,Sun}. In general, these tools
address different {\it features} of non-Markovianity, from the
lack of divisibility of a map~\cite{Huelga} to the ability of the
environment to reciprocate the information transfer from the
system. This process occurs unidirectionally in a Markovian
dynamics~\cite{BreuerMea}, while the re-focusing of information on
the system is the signature of memory effects, as verified in
all-optical set-ups~\cite{natphyscin}. The handiness
of such  instruments has recently triggered the analysis of
non-Markovianity in quantum many-body systems such as quantum spin chains~\cite{tony} or
impurity-embedded ultra-cold atomic systems~\cite{Gabriele} and in
excitation-transfer processes in photosynthetic
complexes~\cite{Rebentrost}. While these studies relate
non-Markovian features to the critical behavior of
a quantum many-body system~\cite{tony,hj}, they also provide a
promising arena where the roots for non-Markovianity can be
researched in physically motivated contexts.

In this paper we explore the competition between  two profoundly
different mechanisms in a simple open quantum model that is
relevant for the physics of nitrogen-vacancy centers in
diamonds~\cite{daaggiu} and molecular nanomagnets~\cite{modenesi}.
Specifically, we address the interplay between the dynamics
induced on a two-level system by its coherent interaction with
other (environmental) spins, and the Markovian process describing
the relaxation of the latter. One would expect that, when such
memoryless dissipative coupling determines the shortest dynamical
timescale of the system, Markovianity should emerge
preponderantly, especially as the number of environmental spins
increases. Indeed, one could imagine that a sort of
``Markovianity-mixing'' property would hold as a result of the
increasing difficulty to re-build the coherence of the system when
many decoherence channels are open.  Quite strikingly, we show
that this is not generally true. In order to do this using a
physically relevant model, general enough to encompass the
unexpected features that we would like to highlight, we consider a
spin-star configuration whose peripheral sites are coupled to {\it
rigid} boson environments, assumed to induce a memoryless
dissipative dynamics. While certainly not exhausting the possible
scenarios that can be tackled, our choice is illustrative since
the degree of non-Markovianity (as defined in
Ref.~\cite{BreuerMea}) {\it can actually  increase} with the
number of peripheral spins, while stronger interactions with the
boson baths only affect its rate of growth. The features of the
system at hand are quite complex and a rich {non-Markovianity
phase diagram} emerges, spanning degrees of memory-keeping effects
all the way down to zero values. This can be exploited to
qualitatively modify the character of the dynamics by engineering
its features via accessible control parameters such as the
detuning between the central and the outer spins. In turn, this
opens up the possibility to implement qubit-state preparation
protocols in an open-system scenario that exploits
non-Markovinity, along the lines of Refs.~\cite{Huelgasteady} and
beyond the well-established Markovian dissipative
framework~\cite{Kraus,Verstraete}.

In the following, we first present the model and its solution in
the simplest terms in Sec. \ref{model}, while the microscopic
description and more sophisticated solution method are presented
in the Appendices. We then proceed to the analysis of the
non-Markovianity of the dynamics in Sec. \ref{nonmasec} and Sec.
\ref{timenonma}. Some concluding remarks are given in Sec.
\ref{conclu}.

\section{The model and its solution}
\label{model} 
\begin{figure}[h!]
\centerline{{\bf (a)}\hskip4cm{\bf (b)}}
\includegraphics[width=8.5cm]{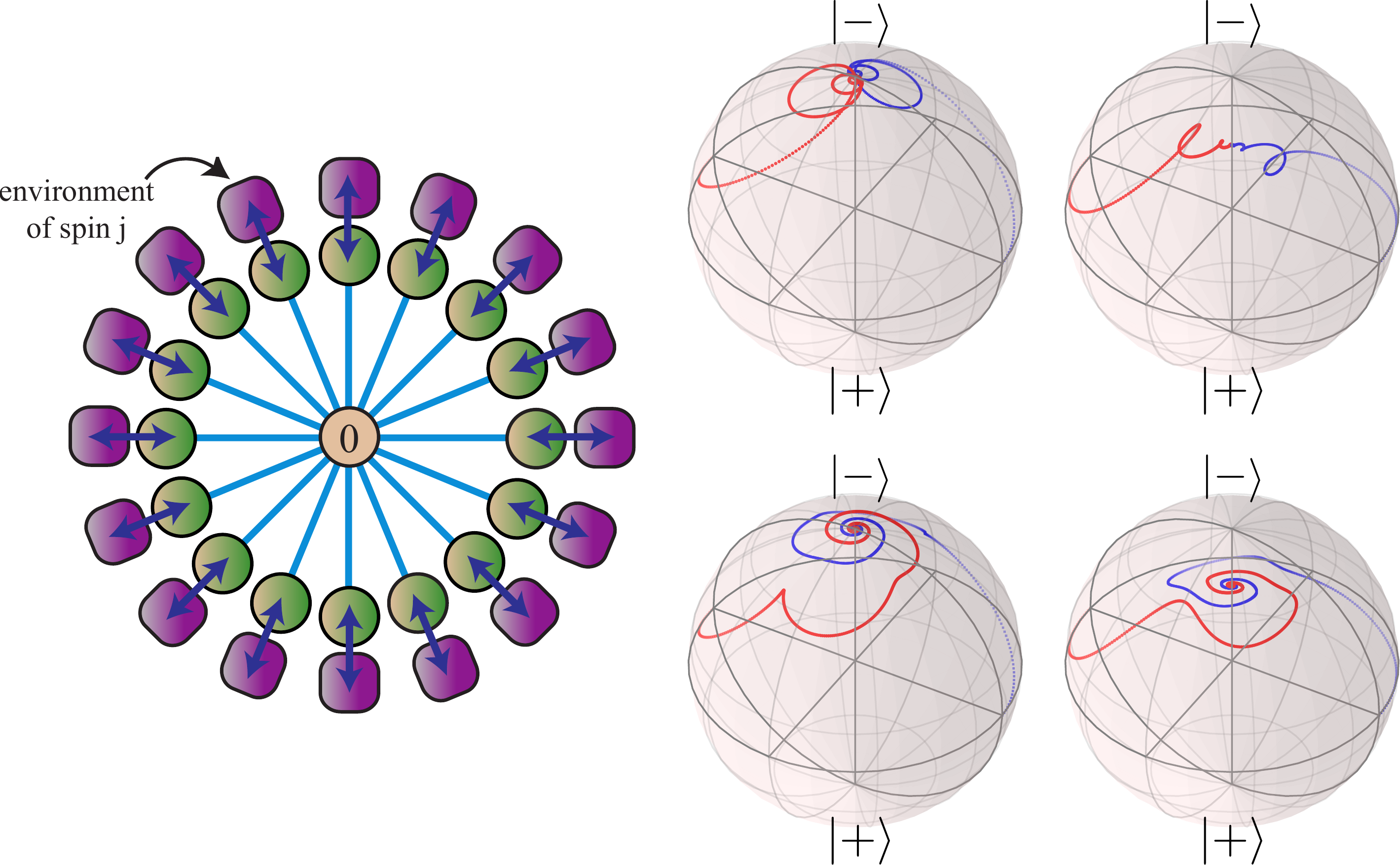}
\caption{{(color on-line): (a)} Scheme of the system: The central spin $0$ interacts with $N$ peripheral spins,
each affected by its own local environment. {(b)} Evolution of
states $\ket{+}_x$
(red trajectory) and $\ket{-}_x$
(blue one) for
a star with $N=4$ peripheral sites. The Bloch spheres in the left
(right) column correspond to the isotropic (anisotropic) spin-spin
coupling. The top (bottom) row is for the resonant (off-resonant
at $\Delta=1/2$) case with $\gamma/J=1$ and $T=0$. In the
isotropic cases, the final state of spin $0$ is pure, while for
$\lambda=\pm 1$ it is mixed. $\Delta\neq0$ prevents the
intersections of the trajectories, which are the dynamical points
at which the trace distance is strictly
null.}\label{bloch_graph}
\end{figure}

The physical set-up that we describe is sketched in
Fig. \ref{bloch_graph} a, which shows a central spin (labelled
$0$) coupled to $N$ outer spins, with bonds along the branch of a
star. Each environmental spin is further coupled to a local boson
reservoir. The evolution of the central spin is ruled by the
master equation
\begin{equation}\label{Eq.lindblad}
\partial_t{\rho_0(t)}=
{\rm Tr}_{S}\{-i[\hat
H,\rho(t)]+\sum_{j=1}^N\hat{\mathcal{L}}_j[\rho(t)]\}
\end{equation}
with $\rho(t)$ the density matrix of the whole system. Each Lindblad superoperator $\hat{\mathcal{L}}_j$ describes
local dissipation at temperature $T$ (the same for
all the baths) as~\cite{Breuer2007}
\begin{equation}
\label{dissipator}
\begin{aligned}
\hat{\mathcal{L}}_j(\rho)&=\gamma(\bar{n}+1)( \hat\sigma^-_j\rho\hat\sigma^+_j -\{\hat\sigma^+_j\hat\sigma^-_j,\rho\}/2)\\
&+\gamma\bar{n}( \hat\sigma^+_j\rho\hat\sigma^-_j -\{\hat\sigma^-_j\hat\sigma^+_j,\rho\}/2),
\end{aligned}
\end{equation}
where $\gamma$ describes the effective coupling of each external
spin to its thermal reservoir, populated by $\bar{n}=(e^{\beta
\omega_j}{-}1)^{-1}$ excitations ($\beta=1/k_b T$, where $k_b$ is
the Boltzmann constant). In what follows, we will consider the
peripheral spins to be initially prepared in
$\otimes^N_{j{=}1}\ket{-}_j$.

To solve the master equation, we use
the damping basis~\cite{Briegel-Englert} made out of tensor products of
eigenoperators of $\hat{\mathcal{L}}_j$. In this
basis, the density matrix of the system reads
\begin{equation}\label{rhodamping}
\rho(t)=\sum_{n=1}^4\sum_{m=1}^{4^N}c_{nm}(t)\hat{\mu}^n_0\otimes \hat{O}_m,
\end{equation}
where
$\hat{\mu}_j^1=(\hat\openone-\frac{1}{2\nbar+1}\hat\sigma^z_j)/2$,
$\hat{\mu}^2_j=\hat\sigma^z_j/2$, $\hat{\mu}^3_j=\hat\sigma^+_j$
and $\hat{\mu}^4_j=\hat\sigma^-_j$ are right eigenoperators of
$\hat{\mathcal{L}}_j$ with eigenvalues
$\lambda^1_j=0,\lambda^2_j=2,\lambda^{3,4}_j=-(2\bar{n}+1)$. The
set of operators $\{\hat{O}_m\}$ is composed of the tensor product
of $N$ damping-basis elements, one for each peripheral spin. Due
to the symmetry of the Hamiltonian, if $\hat{O}_r$ and $\hat{O}_s$
consist of the same elements of the damping bases (although
differing for their order), the respective coefficients must
satisfy $c_{nr}=c_{ns}$. This simple observation allows us to
reduce the number of relevant operators from $4^N$ to
$\tilde{N}\approx N^3$. With the help of the single-spin {dual
damping basis} $\{\check\mu^n_j\}~(n=1,..,4)$, made of left
eigenoperators of $\hat{\cal L}_j$'s, and using the orthogonality
condition ${\rm
Tr}[\hat{\mu}^k_j,\check{\mu}^{k'}_l]=\delta_{kk'}\delta_{jl}$, we
find 
\begin{equation}
\dot c_{rs}(t)=\sum_{n=1}^4\sum_{m=1}^{\tilde{N}}c_{nm}(t)\mathcal{M}_{nmrs}
\end{equation}
with $\mathcal{M}_{nmrs}{=}-i {\rm Tr}\lbrace
(\check{\mu}^r_0{\otimes}\check{O}_s)[\hat
H,\hat{\sigma}^n_0{\otimes}\hat{O}_m]\rbrace{+}\Lambda_m\delta_{rn}\delta_{ms}$
and $\Lambda_m=\sum_{j=1}^N\lambda^m_j$. By calling
$\mathcal{K}(t)=e^{\mathcal{M}t}$, the state of the spin star at
time $t$ is
\begin{equation}
\rho(t){=}
\sum_{r,s,n,m}\mathcal{K}_{nmrs}(t)c_{rs}(0)\hat{\sigma}_r^0\otimes
\hat{O}_s .
\end{equation}
Tracing over the degrees of freedom of the peripheral spins, we
find
\begin{eqnarray}
\rho_0(t) & {=} &
\sum_{r}\left(\sum_{nm}\mathcal{K}_{nmr1}(t)\right)c_{r1}(0)\hat{\sigma}_r^0=
\nonumber \\
&=& \begin{pmatrix}
 \dfrac{\bar{n}}{1+2 \bar{n}}+\dfrac{c_{21}(t)}{2} & c_{31}(t) \\
 c_{31}(t) & \dfrac{(1+\bar{n})}{1+2 \bar{n}}-\dfrac{c_{21}(t)}{2}
\end{pmatrix}.
\label{exactsol}
\end{eqnarray}
This gives the exact solution for the dynamics of the central
spin, valid for any $N$ once the expressions for $c_{rs}(t)$ are
taken. With this at hand, in the next section we evaluate the
amount non-Markovianity of the time evolution.

\section{Non-Markovianity}
\label{nonmasec} To quantify the degree of non-Markovianity of the
dynamical evolution of the central spin described in Eq.
(\ref{exactsol}), we employ the measure put forward in
Ref.~\cite{BreuerMea}, which is based on the idea that memory
effects can be characterized by the information flowing out of the
open system $0$ and quantified in terms of the trace distance
$D[\rho_{0,1}(t),\rho_{0,2}( t)]=\text{Tr} |\rho_{0,1}( t)
-\rho_{0,2}( t)|/2$ between any two of its states
$\rho_{0,j}(t)~(j=1,2)$. The trace distance quantifies the
distinguishability of two states and leads to measure
non-Markovianity as
\begin{equation}
\mathcal{N}=\max_{\rho_{0,j}(0)}\int_{\Omega_+}\partial_t D[\rho_{0,1}(t),\rho_{0,2}(t)],
\label{NonMarkovianity}
\end{equation}%
where $\Omega_+$ is the union of the  intervals where
$\partial_t{D}{>}0$. To provide a general assessment of the
dynamics of spin $0$, we consider the coupling with the external
spins to be described by the anisotropic XY model
\begin{equation} \label{H_S}
\hat H_S= J\sum_{j=1}^N[(1+\lambda){\hat
\sigma_0^x\hat\sigma_j^x}+(1-\lambda){\hat\sigma_0^y\hat\sigma_j^y}],
\end{equation}
where $\lambda$ is an anisotropy parameter and $J$ is the
spin-spin coupling strength. For isotropic coupling
($\lambda{=}0$) and zero temperature, we obtain a simple scaling
law~\cite{breuerscale}: for any $N{>}1$ $\rho_0(t)$ is obtained
from the expression valid for $N=1$ with the re-definition
$J\rightarrow J \sqrt{N}$. This enables the analytic optimization
over the input states entering ${\cal N}$. By calling
$\rho^{kl}_{0,i}{=}\langle k|\rho_{0,i}|l\rangle$, we have
\begin{equation}
D[\rho_{0,1}(t),\rho_{0,2}(t)]{=}\sqrt{\delta \rho^{00}(t)
|g_0(t)|^2{+}\delta \rho^{01}(t) |g_0(t)|}
\end{equation}
where $\delta \rho^{kl}(t)=\rho^{kl}_{0,1}(t)-\rho^{kl}_{0,2}(t)$ and we have introduced
$g_{\nbar}(t)=\exp[{-\frac{1}{2}(G+i\Delta)t}][{(G+i\Delta)\sinh(zt)+z\cosh(zt)}]/{2z}$,
 $z=\sqrt{(G+i\Delta)^2-J^2 N}/2, G={\gamma}(\bar{n}+1/2)$ and
the energy mismatch $\Delta=\epsilon-\epsilon_0$ between the
central and outer spins. The maximum in
Eq.~\eqref{NonMarkovianity} is achieved for the pure states
$\rho_{0,i}=\ket{\psi_i}_0\!\langle\psi_{i}|$ with
$\ket{\psi_i}_0=\cos(\theta_i/2)\ket{-}_0+e^{i\phi_i}\sin(\theta_i/2)\ket{+}_0$.
Here, $(\theta_i,\phi_i)$ are the angles that identify the
respective Bloch vector. ${\cal N}$ is optimized by equatorial
antipodal states  ({\it i.e.} states with $\theta_{1,2}={\pi}/{2}$
and $\phi_{2}= \pi-\phi_1$). In \ref{appB}, we provide an
alternative analytic approach to the evolution of spin $0$ and the
dependence of the trace distance on such angles.

\begin{figure}[h!]
\includegraphics[width=7cm]{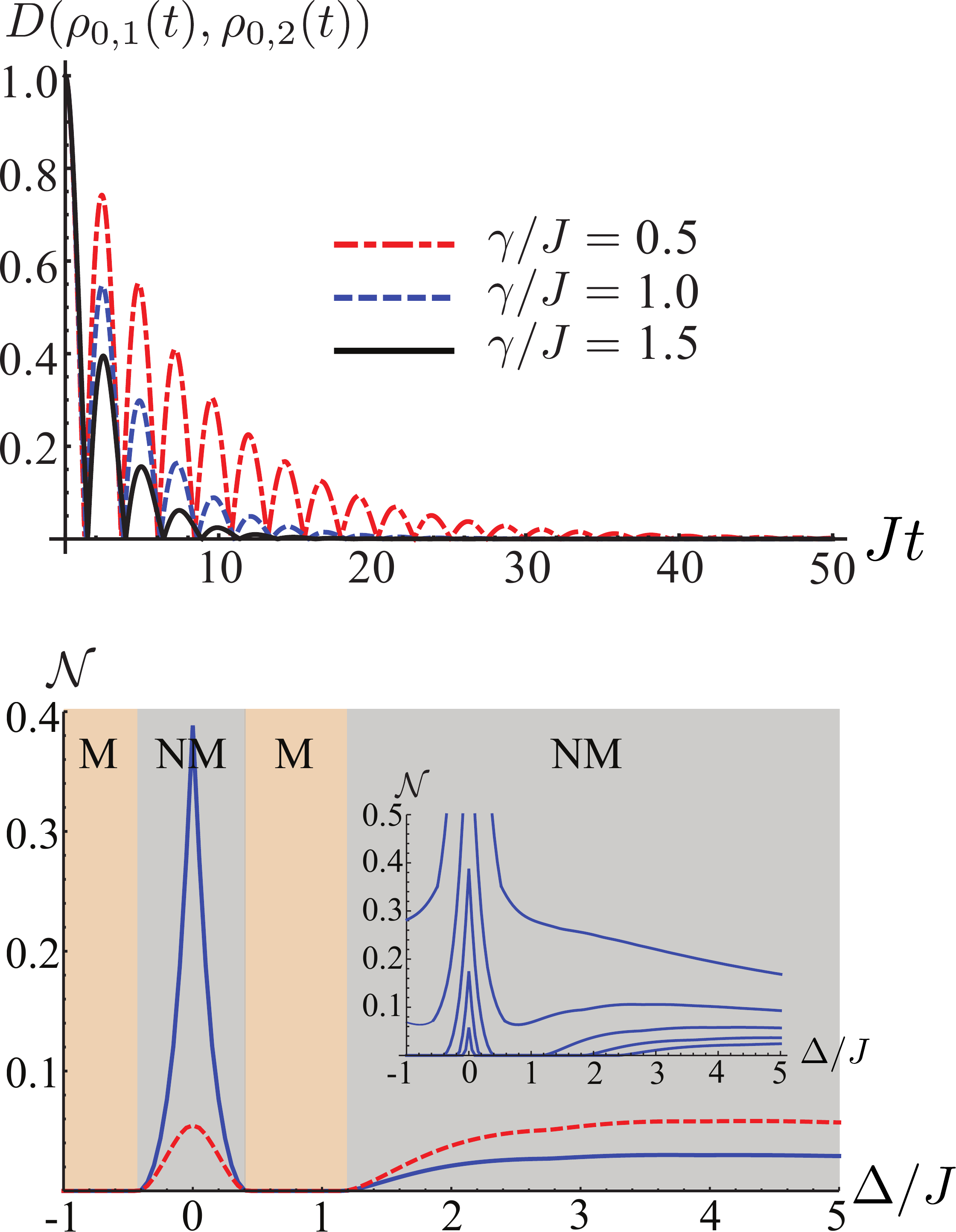}
\caption{(color on-line): (a) Trace distance between the optimal states
$\rho_{0,j}(t)~(j=1,2)$ for $N=6$ peripheral
spins with $\lambda=0$ and $\gamma/J=0.5$ (dot-dashed line),
$\gamma/J=1$ (dashed line) and $\gamma/J=1.5$ (solid line). As the
relaxation time becomes shorter, the revivals of
$D(\rho_{0,1}(t),\rho_{0,2}(t))$ are suppressed as a result of a
reduction of information back-flow from the baths.
(b) ${\cal N}$ against $\Delta$ for
$\gamma/J=\sqrt{N}$. The two lines correspond to
$\theta_{1,2}=\pi/2 \, , (\phi_1, \phi_2)=(0,\pi)$ (solid blue
curve) and $(\theta_1,\theta_2)=(0,\pi)$ (dashed red curve), which
are the optimal states in different detuning regions: ${\cal N}$ is the topmost curve in each
region. There is a finite window of detunings (light-shadowed
region marked as M) where ${\cal N}=0$
[NM marks regions where ${\cal N}\neq0$]. Inset: ${\cal N}$ against $\Delta$ for
$\gamma/J=0.5,0.75,1,1.25$ and 1.5 (from top to bottom curve). }
\label{TRD6}
\end{figure}

The trajectories described on the Bloch sphere by the evolved
states are shown in Fig.~\ref{bloch_graph} (b) [top row, left-most
sphere] where we see that the states tend to intersect, giving
$D=0$. For $\Delta\neq0$, the states that optimize the measure of
non-Markovianity are those with $(\theta_1,\theta_2)=(\pi,0)$ (the
phases being immaterial) as shown in Fig.~\ref{TRD6} (b).
Interestingly, non-zero values of $\Delta$ hinder the
intersections of the state trajectories [cf.
Fig.~\ref{bloch_graph} (b)]. However, this does not prevent the
dynamics to become Markovian at proper working points, as we show
later on.

The evolution of spin $0$ can be characterized using ${\cal N}$.
When the peripheral spins are detached from their respective
baths, any information seeded in the central site undergoes
coherent oscillations from the center to the periphery of the star
and back. For $\lambda=1$ and peripheral spins prepared in
$\openone/N$, the dynamics induced by $\hat{R}\hat H_S\hat{R}$
with $\hat R=\hat\sigma^y_0\otimes^N_{j{=}1}\hat\sigma^y_j$ is
(strongly) non-Markovian at all times~\cite{BreuerMea}. In our
case, the interaction of the outer spins with their environments
radically modifies this picture. As an example, in Fig.~\ref{TRD6}
(a) we plot the trace distance for the optimal states at
$\Delta=0$. We ramp up the spin-bath interaction strength
$\gamma$, at set values of the intra-star coupling $J$, looking
for the influences that an explicitly Markovian mechanism has on
the degree of non-Markovianity that arises from the dynamical
environment to which particle $0$ is exposed. We find a
non-monotonic behavior of the trace distance that results in
non-Markovianity. The quantitative features of $D$ depend on the
actual strength of the Markovian process: as $\gamma$ increases,
the revivals of the trace distance become less pronounced. As
${\cal N}$ depends on the number of temporal regions where
$\partial_tD>0$, Fig.~\ref{TRD6} (a) tells us that ${\cal N}$
decreases as $\gamma$ increases, thus showing that, at resonance,
a strong influence from the rigid environmental baths over the
peripheral spins is sufficient to make the whole process
Markovian.

This is expected as the excitations distributed to the peripheral
spins by spin $0$ find the {\it sink} embodied by the baths. The
reduced ability to feed back information sets ${\cal N}=0$.
However, the general picture is more involved: it is sufficient to
move to the off-resonant case to face a rather rich {\it phase
diagram} of non-Markovianity. Fig.~\ref{TRD6} (b) considers  the
case of coupling mechanisms such that $\gamma/J=\sqrt{N}$ and
explores the effect that an energy mismatch between spin $0$ and
the peripheral sites has on ${\cal N}$. We find two ranges of
values of $\Delta$ for which ${\cal N}=0$, symmetrically with
respect to $\Delta =0$. In between and beyond such regions, ${\cal
N}$ behaves quite distinctively: at resonance, the measure of
non-Markovianity achieves a global maximum (equatorial states
realize the maximum upon which ${\cal N}$ depends). For larger
detunings, ${\cal N}$ changes slowly with $\Delta$ ($\ket{\pm}$
being the optimal states). Clearly, the trend followed by ${\cal
N}$ also depends on $\gamma/J$: small values of $\gamma/J$ push
the dynamics towards strong non-Markovianity, regardless of
$\Delta$, as many coherent oscillations occur between site $0$ and
the periphery before the initial excitation is lost into the
environments. At the same time, the range of detunings for which
${\cal N}=0$ increases with $\gamma$ [cf. inset of Fig.~\ref{TRD6}
(b)]. However non-Markovianity persists, both on and off
resonance, even when $\gamma$ becomes the largest parameter. This
demonstrates an effective control of the degree of
non-Markovianity of the dynamics undergone by spin $0$, which can
be tuned by both the energy mismatch between the outer spins and
the central one, $\Delta$,  and the intra-star coupling strength.,
$J$.
\begin{figure}[h!]\centering
\includegraphics[width=7cm]{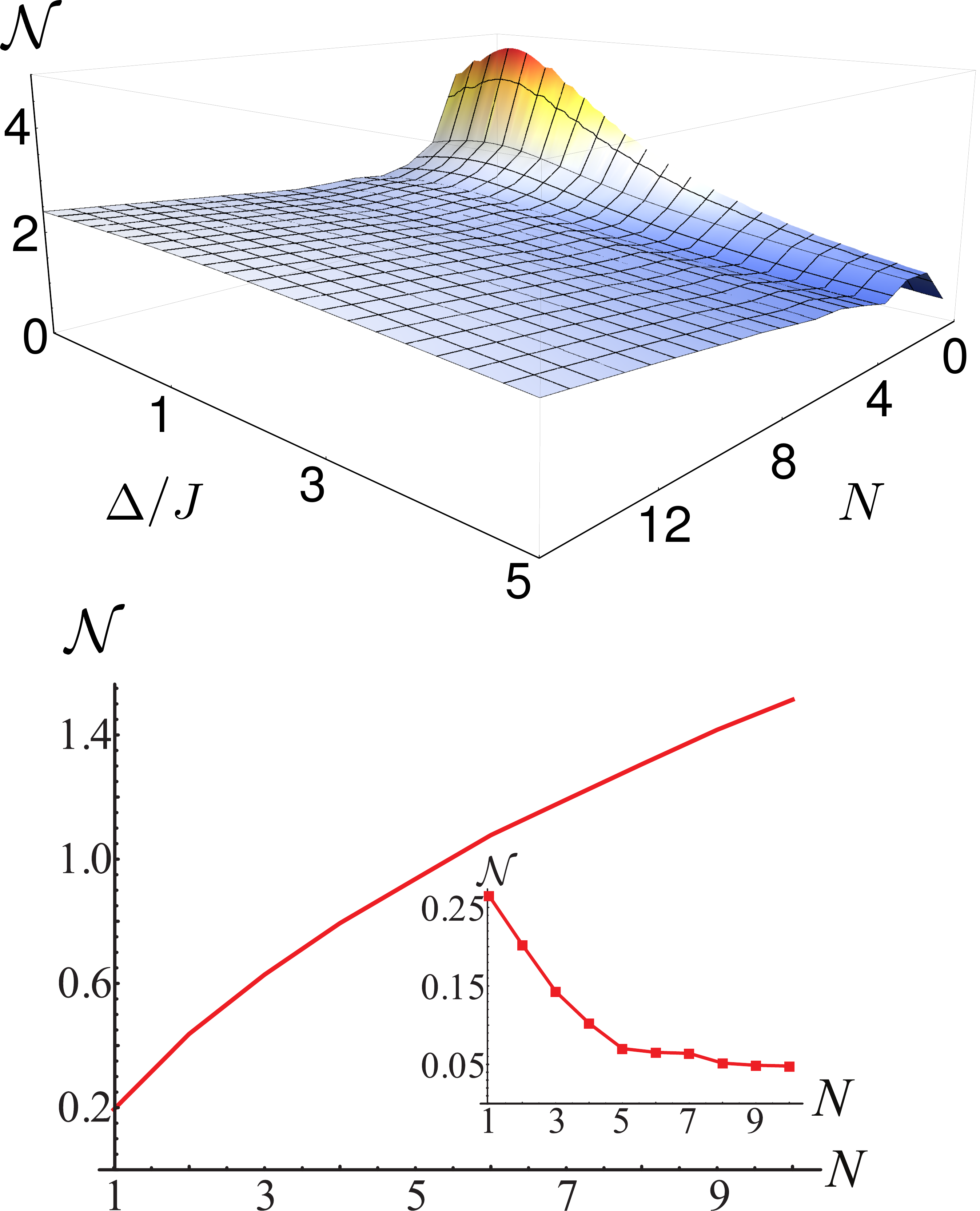}
\caption{(color on-line): (a) ${\cal N}$ against $N$ and $\Delta$ for $\gamma/J=0.1$ and
$\nbar=3$. Differently from $T=0$, except for a small range of
values, the detuning has no effect on the {\it character} of the
dynamics of spin $0$. Strikingly, ${\cal N}$ grows with $N$
(almost linearly for $N\gg1$). {(b)} Analytic behavior of
$\mathcal{N}$ versus $N$ for $\lambda{=}0$, $\Delta=0$, $\gamma=J$
at $T=0$ ($\nbar=0$). Inset: we present the case
corresponding to $\lambda=-1$ [other parameters as in panel (b)].}
\label{NM-ani}
\end{figure}

Our discussions so far were restricted to the isotropic coupling
at zero temperature, $T=\lambda=0$. When the peripheral spins
interact with baths populated by $\nbar$ thermal excitations, the
Markovianity regions disappear. This is seen in Fig.~\ref{NM-ani}
(a) where we show a typical case of the behavior of $\mathcal{N}$
against $\Delta$ and $N$. The anisotropy of the intra-star
coupling is crucial for the determination of the dynamics: for
$\lambda{\neq}0$ the pair of states that maximize ${\cal N}$
changes with the number of peripheral spins. A numerical search
for the optimal states can be performed, leading to quite
surprising results concerning the scaling of ${\cal N}$ with the
size of the spin environment. Intuitively, one would conclude
that, as $N$ grows, the dynamics of spin $0$ will be pushed
towards Markovianity. This is not the case: as shown in
Fig.~\ref{NM-ani} (a), ${\cal N}$ {\it increases} with $N$ if
$\lambda=0$, regardless of $\Delta$. This shows that the
non-Markovian character resists such Markovianity-enforcing
mechanisms and, counter-intuitively, overcomes them. We have
checked this behavior  for  the exact analytical expression
obtained at $\Delta=\lambda=0$ [cf. Fig.~\ref{NM-ani} (b)]. The
picture somehow changes for $\lambda\neq0$: ${\cal N}$ decreases
with the growing dimension of the star. However, even for $N\gg1$
the non-Markovian character is preserved and ${\cal N}$ achieves a
non-null quasi-asymptotic value.

\section{Time development of non-Markovianity}
\label{timenonma} 
\begin{figure}
\includegraphics[width=\linewidth]{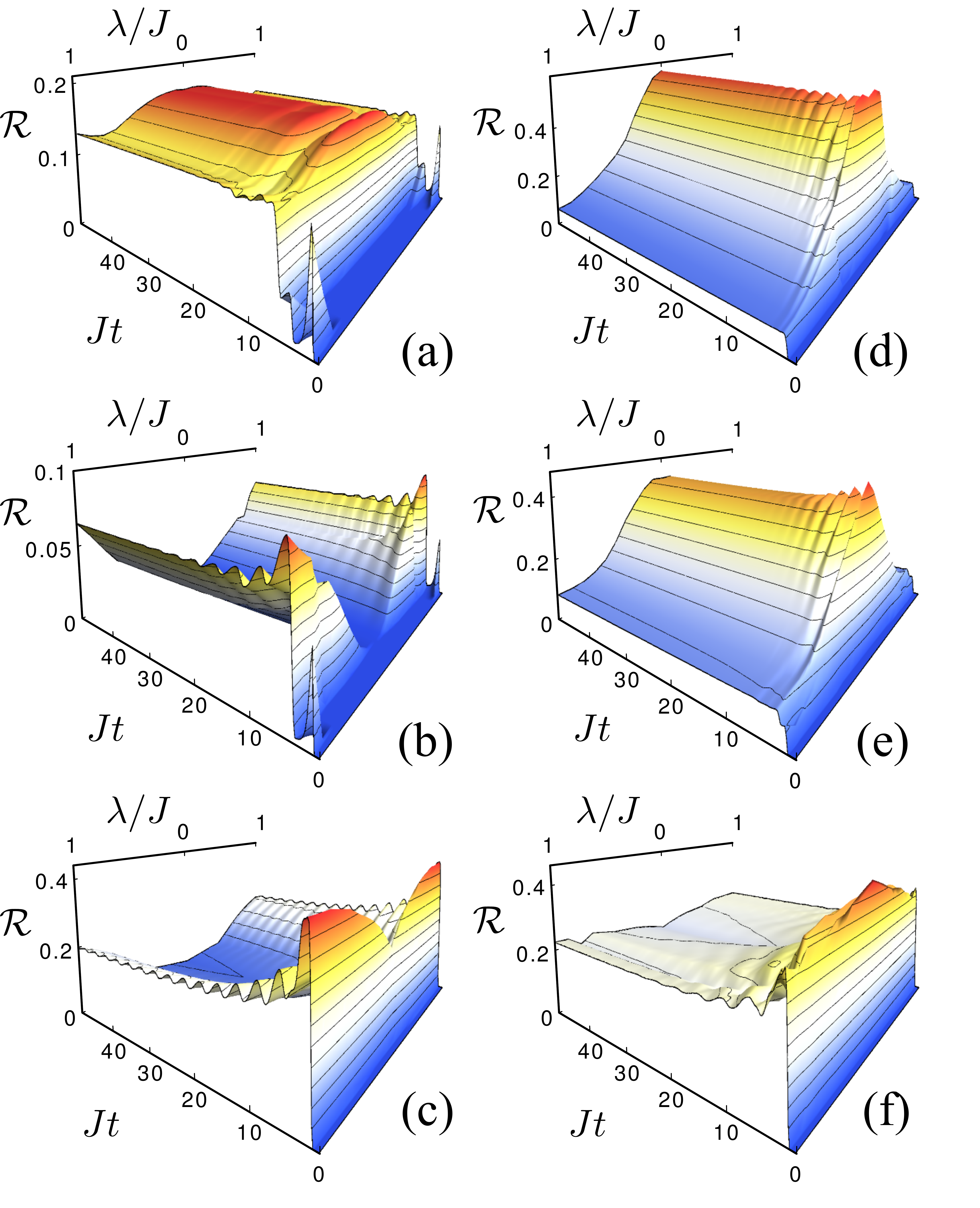}
\caption{(color on-line): Ratio ${\cal R}(t) = \mathcal{N}^+(t)/\mathcal{N}^-(t)$
versus $\lambda$ for a star of $N=8$ sites at $T=0$, with
$\gamma/J=\sqrt{8}$ (left plots) and  $\gamma/J=1$ (right plots)
for three different values of the detunig: $\Delta/J=0$ for the
plots (a) and (d), $\Delta/J=0.7$ for (b) and (e), while
$\Delta/J=3$ for (c) and (f).} \label{seigrafici}
\end{figure}

The non-Markovianity measure gives an integral
characterization of the dynamics. More details on the time
dependence of the system-environment information-exchange process
is obtained by considering the ratio of in-flowing to out-flowing
information, up to a given value $\tau$  of the evolution time. To
this end, we define ${\cal R}(\tau) = \frac{{\cal
N}^+(\tau)}{{\cal N}^-(\tau)}$, where the in-flow [out-flow]
${\cal N}^{+}(\tau)$ [${\cal N}^-(\tau)$] is defined as [minus]
the integral of $\partial_t D$, over the time intervals in which
it is positive (negative), but only up to $\tau$. To evaluate
these quantities explicitly, we chose as input states the same
$\rho_{0,i}$ that optimize the non-Markovianity measure ${\cal N}
\equiv \lim_{\tau\rightarrow \infty} {\cal N}^+(\tau)$.
The ratio ${\cal R}(\tau)$ gives the fraction of the lost
information that returns to the system within $\tau$, and its
behavior is quite different in the various dynamical regimes that
we have identified so far. In Fig.~\ref{seigrafici}, ${\cal
R}(\tau)$ is shown for three values of $\Delta$ corresponding to
the three regions of Fig.~\ref{TRD6} (b). The diverse evolutions
of ${\cal R}(\tau)$ signal qualitatively different dynamical
behaviors of the system, depending on both the detuning and the
anisotropy parameter. At short times, ${\cal R}(\tau)$ is always
zero (information has to flow out of the system before it can come
back), while its first peak is determined by the first revival of
the trace distance [see Fig.~\ref{TRD6} (a)]. Then, its features
become strongly dependent on $\Delta$. At long times and at
resonance, where a maximum of ${\cal N}$ is found for $\lambda=0$,
information oscillates between the star and spin $0$ and ${\cal
R}(\tau)\neq0$ [cf. Fig.~\ref{seigrafici} (a)]. The overall
dynamics is non-Markovian also for the case of Fig.
\ref{seigrafici} (c), where the time behavior of ${\cal  R}(\tau)$
is shown for a large detuning. In this case, however, ${\cal
R}(\tau)$ decays to zero at long times. Thus, the regions of
non-Markovianity in Fig.~\ref{TRD6} (b) correspond to different
behaviors: near resonance, a fraction of information comes back to
the system, different input states remain distinguishable even at
long times and thus no equilibrium state is found. For large
detunings, non-Markovianity is built up at short times, while
different input states converge towards a long-time equilibrium.
On the other hand, for intermediate values of the detuning [{\it
i.e.} for $\Delta$ in the Markovianity region of Fig.~\ref{TRD6}
(b)] and $\lambda=0$, there is no back-flow.
Even for
$\lambda\neq0$, the fraction of information that comes back is
quite small. The picture changes when $J$ increases, the evolution
becoming increasingly non-Markovian and the role of the anisotropy
being fully reversed: $\lambda=0$ implies a larger ${\cal
R}(\tau)$, persisting for longer times at resonance.
\section{Concluding remarks}
\label{conclu} We have used a measure of non-Markovianity to show
the possibility to control the dynamics of an open quantum system
coupled to many independent decohering channels. We have
highlighted the key role played by the detuning and the degree of
anisotropy of the system-environment coupling: both can be used to
explore a rich non-Markovianity phase diagram, where qualitatively
different scaling laws with the number of decoherence channels are
found. The ability to switch from a Markovian to a non-Markovian
regime by means of a local parameter could be used to prepare a
quantum system in a desired state: indeed the Markovian character
of processes can be employed for state engineering and information
manipulation \cite{Kraus,Verstraete}. On other hand, while the
formation of a steady entangled state is supported by
non-Markovianity, a purely Markovian dynamics produces separable
steady states \cite{Huelgasteady}.

\section*{Aknowledgement} We acknowledge financial support from the UK
EPSRC (EP/G004759/1). SL thanks the Centre for Theoretical Atomic,
Molecular and Optical Physics for hospitality during the early
stages of this work.

\section{Appendix A}
\label{appA}
The total Hamiltonian of the spin-star system $\hat H{=}\hat
H_0{+}\hat H_S{+}\hat H_B$ consists of a few contributions. The
first one is the system's free energy [we take units such that $\hbar{=}1$ throughout the
paper] $\hat H_0 = \sum_{j=0}^N{\epsilon_j \hat
\sigma^{z}_{j}}+\sum_{j=1}^N{\sum_k{\omega_k  \hat b_{k,j}^\dag
\hat b_{k,j}}}$, describing the free evolution of $N+1$ spin-$1/2$
particles [here, $\hat\sigma^s_j$ is the $s$-Pauli matrix of
spin $j$ ($s=x,y,z$)], each with transition frequency $\epsilon_j$ between
spin states $|-\rangle_j$ and $|+\rangle_j$. The second term in
$\hat H_0$ describes the energy of $N$ sets of $M_j$ harmonic
modes (one set per peripheral spin of the star) with creation
(annihilation) operators $\hat b_{k,j}^{\dagger}$ ($\hat b_{k,j}$)
which satisfy the commutation relations $[\hat b_{k,j},\hat
b_{k',l}^{\dagger}] =\delta_{kk'}\delta_{jl}$. The central and
peripheral spins are coupled by  $\hat H_S$, whose explicit form
will be specified later on. Each peripheral spin
interacts with its own bath as $\hat
H_B=\sum^N_{j{=}1}\hat{H}_{B,j}$, where
\begin{equation} \label{H_B}
  \hat H_{B,j}= \sum_k  (g_{k,j} \hat \sigma^+_j \hat b_{k,j} + g^*_{k,j} \hat \sigma^-_j \hat b^{\dagger}_{k,j})
\end{equation}
with $\hat\sigma^\pm_j=(\hat\sigma^x_j\pm i\hat\sigma^y_j)/2$. We
assume that the their local bath induce a Markovian dynamics of
the peripheral spins and take uniform couplings, so that the
evolution of the central spin is ruled by
\begin{equation}\label{Eq.lindblad}
\partial_t{\rho_0(t)}=
{\rm Tr}_{S}\{-i[\hat H,\rho(t)]+\sum_{j=1}^N\hat{\mathcal{L}}_j[\rho(t)]\}
\end{equation}

\section{Appendix B}
\label{appB}
Here we provide an alternative solution to the dynamics of the system.
In the interaction picture with respect to $\hat H_0$ the Schr\"odinger equation reads
\begin{equation} \label{Schr}
 \partial_t|\Psi(t)\rangle = -i\hat H_I(t)|\Psi(t)\rangle,
\end{equation}
where the interaction Hamiltonian is given by
\begin{eqnarray}
\label{HIt}
 H_I(t) = J\sum_{j=1}^N(\sigma_0^+(t){\sigma_j^-(t)}+\sigma_0^-(t){\sigma_j^+(t)})\nonumber\\
+ \sum_{j=1}^N{\sum_k ( g_k \sigma^j_+(t) b_k(t) + g^*_k \sigma^j_-(t) b^{\dagger}_k(t))}
\end{eqnarray}
with
\begin{equation}
\begin{cases}
 \sigma_j^{\pm}(t) = \sigma_j^{\pm} e^{\pm i \epsilon_j t} \;\;\;\;(j=0,...N)\\
  b_k(t)=g_k b_k e^{-i \omega_k t}\\
  b_k^\dag(t)=g_k b_k^\dag e^{+i \omega_k t}.
\end{cases}
\end{equation}
The operator $N=\sum_j({\sigma_j^z} +(\sum_k b_k^{\dagger}b_k)_j)$
counts the number of excitations in the system and commutes with
the total Hamiltonian $H$, so that any initial state of the form
\begin{eqnarray} \label{statotempo0}
 |\Psi(0)\rangle = (c_0|-\rangle^0+c_1(0)|+\rangle^0) \mathbf{|0\rangle}^S\mathbf{|0\rangle}^B+\nonumber\\
 \sum_{j=1}^N c_j(0) |-\rangle^0\mathbf{|j\rangle}^S\mathbf{|0\rangle}^B +\sum_{j=1}^N \sum_k c_{kj}(0)|-\rangle^0\mathbf{|0\rangle}^S\mathbf{|k\rangle}_j^B \nonumber
\end{eqnarray}
evolves after time $t$ into the state
\begin{eqnarray} \label{statotempot}
 |\Psi(t)\rangle = (c_0|-\rangle^0+c_1(0)|+\rangle^0)\mathbf{|0\rangle}^S\mathbf{|0\rangle}^B+\nonumber\\
 \sum_{j=1}^N c_j(t) |-\rangle^0\mathbf{|j\rangle}^S\mathbf{|0\rangle}^B +\sum_{j=1}^N \sum_k c_{kj}(t)|-\rangle^0\mathbf{|0\rangle}^S\mathbf{|k\rangle}_j^B
\end{eqnarray}
where the state $\mathbf{|0\rangle}^S$ denotes the product state
$\otimes_{j=1}^N|-\rangle^j$ and
$\mathbf{|j\rangle}^S=\sigma^+_j\mathbf{|0\rangle}^S$ for the
sites on the star; $\mathbf{|0\rangle}^B$ is the vacuum state of
all the reservoirs, and $\mathbf{|k\rangle}_j^B  =
b_k^\dagger|0\rangle_j$ the state with one particle in mode $k$ in
the $j$th reservoir.\\
The amplitude $c_0$ is constant in time because of $H_I(t)|-\rangle^0 \mathbf{|0\rangle}^S \mathbf{|0\rangle}^B = 0$.\\
Substituting Eq.~\eqref{statotempot} into the Schr\"odinger equation \eqref{Schr} one finds
\begin{eqnarray}
 &\frac{d}{dt}c_1(t)=&-i J \sum_{j=1}^N c_j(t) e^{i(\epsilon_0-\epsilon_j)t},
 \label{eq-1}\nonumber\\
 &\frac{d}{dt}c_j(t)=&-i J c_1(t)e^{-i(\epsilon_0-\epsilon_j)t}-i\sum_k c_{kj}(t) g_{kj} e^{i(\epsilon_j-\omega_{kj})t}
 \label{eq-2}\nonumber\\
 &\frac{d}{dt}c_{kj}(t)=&-i g_{kj}^* c_j(t) e^{-i(\epsilon_i-\omega_{kj})t},
 \label{eq-3}
\end{eqnarray}
We assume in the following that $c_j(0)=c_{kj}(0)=0$. This means that the
two level systems on the star are in the $|-\rangle$ state and that each environment is in the vacuum state initially.\\
The total initial state is given by the product state
\begin{equation} \label{statoiniziale}
  |\Psi(0)\rangle = \big(c_0|-\rangle^0 + c_1(0)|+\rangle^0\big)\mathbf{|0\rangle}^B\mathbf{|0\rangle}^S.
\end{equation}
Formally integrating Eq.~\eqref{eq-3} and substituting into Eq.~\eqref{eq-2}
one obtains the system for the amplitude $c_1(t),c_j(t)$,
\begin{equation}\begin{cases}\begin{aligned}
&\frac{d}{dt}c_1(t)=&&-i J \sum_{j=1}^N c_j(t) e^{i(\epsilon_0-\epsilon_j)t}&\\
&\frac{d}{dt}c_j(t)=&&-i J c_1(t)e^{-i(\epsilon_0-\epsilon_j)t}-&\\
&&&\int_0^t c_j(t_1) \sum_k |g_{kj}|^2 e^{i(\epsilon_j-\omega_{kj})(t-t_1 )}dt_1&
 \end{aligned}
\end{cases} \label{eq-21}\end{equation}
We can define the kernels $f_j(t-t_1)$ describing the two-point
correlation function of each reservoir, which are the Fourier
transform of the respective environmental spectral density
\begin{equation}
\label{kernel}
 f_j(t-t_1) = 
 \sum_k |g_{kj}|^2 e^{i(\epsilon_j-\omega_{kj})(t-t_1)}.
\end{equation}
For the moment, we do not make any restrictive hypothesis on the
form of $f_j$, so that our results will be valid for an
environment with a generic spectral density. In order to solve the
system above it is convenient to pass in the Laplace domain:
\begin{equation}\begin{cases}\begin{aligned}
 &s \tilde{c_1}[s]&=&\;c_1(0)-i J \sum_{j=1}^N \tilde{c_j}[s+i(\epsilon_0-\epsilon_j)]\\
 &s \tilde{c_j}[s]&=&-i J \tilde{c_1}[s-i(\epsilon_0-\epsilon_j)]- \tilde{c_j}[s]\tilde{f}_j[s]
 \end{aligned}\end{cases} \label{eq-22}\end{equation}
Solving the second of Eq.~\ref{eq-22} respect to
$\tilde{c_j}[s]$,assuming that all the reservoirs are the same
($f_j(t)=f(t)~~\forall~j$), and substituting in the first we get
\begin{equation}
\tilde{c_1}[s]=c_1(0)\dfrac{s-i\Delta-f[s-i(\epsilon_0-\epsilon)]}{s^2-is(\epsilon_0-\epsilon)-i s f[s-i(\epsilon_0-\epsilon)]+J^2 N}\label{c1s}\nonumber
 \end{equation}
where $\Delta=(\epsilon_0-\epsilon)$ ($\epsilon_j=\epsilon\;\;\forall\;j$).

To specify the model, but still retaining a general enough
description, we consider a Lorentzian spectral density for each
bath (which gives rise to an exponentially decaying correlation
function):
 \begin{equation}
J(\omega ) = \frac{1}{{2\pi }}\frac{{\gamma \lambda ^2 }}{{(\epsilon_j-\delta  - \omega )^2  + \lambda ^2 }}.
\end{equation}
Here $\delta=\epsilon_j-\omega_c$ is the detuning of the center
frequency of the bath $\omega_c$ and the frequency of the
two-level system $\epsilon_j$, the parameter $\lambda$ defines the
spectral width of the environment, which is associated with the
reservoir correlation time by the relation $\tau_B=\lambda^{-1}$
and the parameter $\gamma$ is related to the relaxation time scale
$\tau_R$ by the relation $\tau_R=\gamma^{-1}$.\\
We will consider $\delta=0$, and in this case we may distinguish
between the Markovian and the non-Markovian regimes (for the
dynamics of the environmental spins themselves) using the ratio of
$\gamma$ and $\lambda$:
$\gamma<\frac{\lambda}{2}$ gives a Markovian regime and $\gamma>\frac{\lambda}{2}$
corresponds to non-Markovian regime.\\
Substituting in Eq.~ \ref{c1s} and anti-transforming we have
$c_1(t)=c_1(0)G(t)$ with
\begin{equation}
\begin{aligned}
G(t)=c_1(0)\frac{\sum^3_{i=1} (-1)^{i-1}e^{t\alpha_i}(\alpha_j{-}\alpha_k)[\delta^2_i{+}(\delta_i{+}\gamma/2)\lambda]}{(\alpha_1-\alpha_2)(\alpha_2-\alpha_3)(\alpha_1-\alpha_3)},
\end{aligned}\label{Gt}
\end{equation}
where $\delta_i=\alpha_i{-}i\Delta$, $j,k=1,2,3$ and for $j{<}k$. Here, $\alpha_i$'s are the roots of the equation
\begin{align}
p(s)=&s^3+(2 i\Delta+\lambda)s^2+&\\ \nonumber
&(J^2N+\Delta^2+i\Delta\lambda+\lambda\gamma/2)s+J^2N(i\Delta+\lambda)&.\nonumber
\end{align}
Already at this point, it is evident how the only effects of
increasing $N$ is to redefine the coupling constant $J$.

The solution of the Schr\"odinger equation of the total system with initial states of the form \eqref{statoiniziale} lyes in the sector of the Hilbert space corresponding to zero or one excitations.\\
We can construct the exact dynamical map describing the time-evolution of the reduced density matrix of the central spin which is given by
\begin{equation} \label{RHO-QUBIT}
 \rho(t) = {\rm Tr}_{S+B} \{ |\Psi(t) \rangle \langle \Psi(t)| \} =
 \left(\begin{array}{cc}
  \rho_{11}(t) & \rho_{10}(t) \\
  \rho_{01}(t) & \rho_{00}(t)
  \end{array}\right),
\end{equation}
where $\rho_{ij}(t)=\langle i|\rho(t)|j\rangle$ for $i,j=0,1$.
Using Eq.~\eqref{statotempot} and Eq.~\eqref{Gt} we find
\begin{eqnarray}
 \rho_{11}(t) &=& 1-\rho_{00}(t) = |c_1(0)G(t)|^2, \\
 \rho_{10}(t) &=& \rho_{01}^*(t) = c_0^*c_1(0)G(t).
\end{eqnarray}
The optimization of the initial states in Eq. (\ref{NonMarkovianity}) obtains
the maximally possible non-Markovianity of a particular quantum evolution.\\
In our case, the maximization is achieved by pure states, thus we
choose as initial states for Eq.~\ref{statoiniziale}
\begin{eqnarray}
|\Psi_1(0)\rangle = \big(\cos(\theta_1)|-\rangle +e^{i\phi_1}\sin(\theta_1)|+\rangle\big)\mathbf{|0\rangle}_B\mathbf{|0\rangle}_S.\\
|\Psi_2(0)\rangle = \big(\cos(\theta_2)|-\rangle + e^{i\phi_2}\sin(\theta_2)|+\rangle\big)\mathbf{|0\rangle}_B\mathbf{|0\rangle}_S.
\end{eqnarray}
With these states, the trace distance takes the form
\begin{equation}
\frac{1}{2} |G(t)| \sqrt{|G(t)|^2 (\cos (\theta_1)-\cos (\theta_2))^2+(\sin(\theta_1)+\sin(\theta_2))^2}
\end{equation}
where we used the fact that, since $H$ is invariant under rotations along z-axis, the maximum is obtained for $\phi_1-\phi_2=\pi$.\\

\end{document}